%% file: paper.tex
\documentclass[runningheads]{llncs}
\usepackage[utf8]{inputenc}

\usepackage{hyperref}
\usepackage{longtable}
\usepackage{subcaption}
\usepackage{graphicx}

\usepackage{listings}
\usepackage[dvipsnames]{xcolor}
\usepackage{textcomp}
\usepackage{lstcoq}
\usepackage{supertabular}
\usepackage{amsmath,amssymb}
\usepackage{wrapfig}

\newcommand{\todo}[1]{}
\input{michelsonlang}
\input{albertlang}

\begin{document}
\title{Making Tezos smart contracts more reliable with Coq}
\author{Bruno Bernardo \and
Rapha\"el Cauderlier \and
Guillaume Claret \and
Arvid Jakobsson \and
Basile Pesin \and
Julien Tesson}
\authorrunning{B. Bernardo \and R. Cauderlier \and G. Claret \and A.
  Jakobsson \and B. Pesin \and J. Tesson}
\institute{Nomadic Labs, Paris, France\\ \email{\{first\_name.last\_name\}@nomadic-labs.com}}
\maketitle              % typeset the header of the contribution
%

% from Albert
% \begin{abstract}
%   Tezos is a smart-contract blockchain. Tezos smart contracts are
%   written in a low-level stack-based language called Michelson.
%   In this article we present Albert, an intermediate language for Tezos
%   smart contracts which abstracts Michelson stacks as linearly typed
%   records. We also describe its compiler to Michelson, written in Coq,
%   that targets Mi-Cho-Coq, a formal specification of Michelson
%   implemented in Coq.
% \end{abstract}

\begin{abstract}
  Tezos is a smart-contract blockchain. Tezos smart contracts are
  written in a low-level stack-based language called Michelson. This
  article gives an overview of efforts using the Coq proof assistant to
  have stronger guarantees on Michelson smart contracts: the
  Mi-Cho-Coq framework, a Coq library defining formal semantics of
  Michelson, as well as an interpreter, a simple optimiser and a
  weakest-precondition calculus to reason about Michelson smart
  contracts; Albert, an intermediate language that abstracts Michelson
  stacks with a compiler written in Coq that targets Mi-Cho-Coq.% ;
  % finally we also present our experimentations to link the Michelson
  % interpreter from a Tezos node written in OCaml to the Mi-Cho-Coq
  % implementation in Coq.
\end{abstract}
% \begin{abstract}
% Michelson - Mi-Cho-Coq - coq-of-ocaml and Michelson interpreter and typechecker - Albert
% \end{abstract}

The final authenticated publication is
available online at
\url{https://doi.org/10.1007/978-3-030-61467-6_5}.

\section{Introduction}
%\input{intro}
%\todo{use style for Michelson and Albert}

Tezos~\cite{goodman2014positionpaper,goodman2014whitepaper,tezosIntro2019ABT}
is a public blockchain launched in June 2018. An open-source
implementation of a Tezos node in OCaml is
available~\cite{tezosGitLab}. Tezos has smart-contracts capabilities,
a proof-of-stake consensus algorithm, and a voting mechanism that
allows token holders to vote for changes to a subset of the codebase
called the \emph{economic protocol}.
This subset contains, amongst other elements, the voting rules
themselves, the consensus algorithm, and the interpreter for
Michelson, the language for Tezos smart contracts.

%The language of the smart contracts stored in the Tezos blockchain is
Michelson~\cite{michelsonwhitedoc} is a stack-based
Turing-complete domain-specific language with a mix of low-level and
high-level features. Low-level features include stack manipulation
instructions. High-level features are high-level data types (option
types, sum types, product types, lists, sets, maps, and anonymous
functions) as well as corresponding instructions.
Michelson is strongly typed: data, stacks and instructions have a
type. Intuitively the type of a stack is a list of the types of its
values, and the type of an instruction is a function type from the input
stack to the output stack.
% example the type of the instruction that adds two integers is:
% and the Michelson interpreter checks that
% instructions are applied on properly formed stacks. (E.g. the `ADD`
% instructions, which adds two integers, must be applied on a stack with
% at least two elements and where the top two elements are integers.)
The combination of high and low level features is the result of a
trade-off between the need to meter resource consumption (computation
{\it gas} and storage costs) and the willingness to have strong
guarantees on the Michelson programs.

Michelson has been designed with formal verification in mind: its
strong type system guarantees that there can be no runtime error apart
from explicit failure and {\it gas} or token exhaustion. Furthermore,
its OCaml implementation uses GADTs~\cite{xi2003gadts} which gives
subject reduction for free.
%note: GADT interpreter in OCaml?
In this article, we describe  a couple of efforts using Coq to make Tezos
smart contracts more reliable.
The first one is Mi-Cho-Coq, a Coq library that
implements a Michelson interpreter, a weakest
precondition calculus enabling the functional verification of
Michelson programs as well as a very simple optimiser.
%
% Furthermore, there is a Coq implementation of a Michelson interpreter,
% called Mi-Cho-Coq~\cite{michocoq} and the functional correctness of
% some contracts have been formally verified using this
% framework~\cite{multisigMiChoCoq,multisigArthur}.

The second one is Albert, an intermediate language that abstracts
Michelson stacks into records with named variables, for which we have
implemented a compiler in Coq that targets Mi-Cho-Coq.
Because of its low-level aspects, it is hard to write
Michelson programs, and as a consequence, higher-level languages
compiling to Michelson, such as LIGO~\cite{ligo} or
SmartPy~\cite{smartpy} have been developed in the Tezos ecosystem.
Ideally, there would be certified compilers from these high-level
languages to Michelson, and formal proofs of smart-contracts would be
done directly at the higher level and not at the Michelson level, as
it is done with Mi-Cho-Coq.
The goal of Albert is to facilitate the implementation of certified
compilers to Michelson by being used as a target for
certified compilers from high-level languages.

This article is organised as follows: in
Section~\ref{sec:michelson-vote} we illustrate the Michelson language
with an example, we describe the Mi-Cho-Coq library in
Section~\ref{sec:michocoq} and Albert in Section~\ref{sec:albert}.
Future and related work is discussed in Section~\ref{sec:future}.
% Finally, we describe current efforts to formally prove the equivalence
% between the OCaml Michelson interpreter of a Tezos node and the
% Mi-Cho-Coq interpreter.

%\section{Michelson}

\section{Example of a Michelson contract}
\label{sec:michelson-vote}
The goal of this section is to give the reader an intuitive feeling
of the Michelson language. Our explanations will be illustrated by an
example of a Michelson program, a voting contract, presented in
figure~\ref{fig:vote}.
This contract allows any voter to vote for a predefined set of choices.
Voting requires a fee of 5 tez. It is possible to vote multiple times.
The predefined set of choices is the
initial storage chosen at the deployment of the contract. In this
example, we assume that we want to vote for our favourite proof
assistant amongst Agda, Coq and Isabelle (cf.
fig.~\ref{fig:vote-storage}). Initially, each tool has obviously 0 vote.

Smart contracts are accounts that can contain code and storage. Calls
to a smart-contract provokes the execution of the code contained in
the account with the data sent during the transaction as input
arguments for the code. The execution of the code can lead to a
modification of the storage, it can also generate other transactions.
The storage must be initialised when the contract is deployed on the
chain.

A Michelson program is thus composed of three fields: \verb+storage+,
\verb+parameter+ and \verb+code+ that respectively contain types of
the {\em storage} of the account, of the {\em parameter} that is sent
during the transaction, or the {\em code} contained in the account.

In the case of the voting contract, the \verb+storage+ (l.1) is a map
from strings to integers: the keys are the different choices for the
vote, the values are the number of votes for a given choice. The
\verb+parameter+ (l.2) is a string that represents the ballot that has
been chosen by the caller of the contract.

As mentioned in the introduction, Michelson is a stack based language.
The calling
convention of Michelson is the following: the initial stack contains
one element that is a pair of which the left member is the parameter
sent by the transaction and the right element is the initial storage.
At the end of the execution of a Michelson script, the stack must have
one element that is a pair containing on the left a list of operations
(\emph{e.g.} a transaction) that will be executed afterwards and on
the right the updated storage.
Each instruction takes an input stack and rewrites it into an output
stack.
In the comments of the example are written the content of the stack
before and after the execution of some groups of instructions.
The program starts by verifying that enough tokens were sent by the
voters. This is implemented in lines 5 to 8. The amount sent by the
voter is pushed to the stack (?AMOUNT?) followed by the minimum amount
required ($5000000$ $\mu tez$ \emph{i.e} 5 $tez$). ?COMPARE? pops the
two amounts and pushes 1,0,-1 whether the 5 $tez$ threshold is
greater, equal or smaller than the amount sent by the voter.
If the threshold is greater then the contract will ?FAIL?: the
execution of the contract is stopped and the transaction is cancelled.
\todo{clarify that the tokens sent are not refunded}
Lines 9 to 11 contain stack manipulations that duplicate (with
instruction ?DUP?) the ballot and the current vote count. ?DIP {code}?
protects the top of the stack by executing ?code? on the stack without
its top element.
The next block, from line 12 to 18, tries to ?UPDATE? (l.18) the map with an
incremented number of votes for the chosen candidate. This only
happens if a candidate is a valid one, that is, if it is equal to one
of the keys of the map.
Indeed, in line 13 ?GET? tries to retrieve the number of votes: it
returns ?None? if the chosen candidate is not in the list or ?Some i?
if the candidate is in the list and has ?i? votes. ?ASSERT_SOME? will
fail if ?None? is at the top of the stack and will pop ?Some i? and
push ?i? at the top.
The incrementation by one of the number of votes for the chosen
candidate is done in l.15.\todo{conclusion to the section. some high
  level remarks. maybe links to other docs: michelson tezos dev
  reference, michelson nl reference or try michelson}

\newbox\voting
\begin{lrbox}{\voting}
\begin{lstlisting}[numbers=left]
storage (map string int); # candidates
parameter string; # chosen
code {
  # (chosen, candidates):[]
  AMOUNT;  # amount:(chosen, candidates):[]
  PUSH mutez 5000000; COMPARE; GT;
  # (5 tez > amount):(chosen, candidates):[]
  IF { FAIL } {}; # (chosen, candidates):[]
  DUP; DIP { CDR; DUP };
  # (chosen, candidates):candidates:candidates:[]
  CAR; DUP; # chosen:chosen:candidates:candidates:[]
  DIP { # chosen:candidates:candidates:[]
        GET; ASSERT_SOME;
        # candidates[chosen]:candidates:[]
        PUSH int 1; ADD; SOME
        # (Some (candidates[chosen]+1)):candidates:[]
      }; # chosen:(Some (candidates[chosen]+1)):candidates:[]
  UPDATE; # candidates':[]
  NIL operation; PAIR # (nil, candidates'):[]
}
\end{lstlisting}
\end{lrbox}
\newbox\votingstorage
\begin{lrbox}{\votingstorage}
\begin{lstlisting}
{Elt "Agda" 0 ; Elt "Coq" 0 ; Elt "Isabelle" 0}
\end{lstlisting}
\end{lrbox}
\begin{figure}[h!]
\centering
\captionsetup[subfigure]{position=b}
  \subcaptionbox
    {\label{fig:vote}}
    {\usebox\voting}
    \hfill
\centering
  \subcaptionbox
    {\label{fig:vote-storage}}
    {\usebox\votingstorage}
  % \begin{subfigure}{0.3\textwidth}
  %   \usebox\votingstorage
  %   \caption{\label{fig:vote-storage}}
  % \end{subfigure}
    \caption{A simple voting contract \subref{fig:vote} and an example of initial storage \subref{fig:vote-storage}}
\end{figure}

\section{Mi-Cho-Coq: defining clear semantics of Michelson}
\label{sec:michocoq}

Mi-Cho-Coq~\cite{michocoq} is a Coq library that contains an
implementation of the Michelson syntax and semantics as well as a
weakest precondition calculus that facilitates functioning
verification of Tezos smart contracts. Also, we have recently
implemented a certified optimiser that performs basic simplifications
of Michelson programs.

Mi-Cho-Coq has already been presented in~\cite{michocoq} and we refer
the reader to this publication for more details. Here we present
Mi-Cho-Coq more succintly and focus on high-level additions and changes, as
the implementation has evolved significantly.

\subsection{Syntax, Typing and Semantics of Michelson in Coq}

\subsubsection{Syntax and typing}
The data stored in the stacks have a type defined in the \coqe{type}
inductive. The type of a stack is a list of \coqe{type}.
%
% \begin{lstlisting}
% Inductive comparable_type : Set :=
% | string | nat | int | bytes | bool | mutez | address | key_hash | timestamp.
%
% Inductive comparable_type : Set :=
% | Comparable_type_simple : simple_comparable_type -> comparable_type
% | Cpair : simple_comparable_type -> comparable_type -> comparable_type.
%
% Inductive type : Set :=
% | Comparable_type (_ : simple_comparable_type)
% | key | unit | signature | option (a : type) | list (a : type)
% | set (a : comparable_type) | contract (a : type)
% | operation | pair (a : type) (b : type)
% | or (a : type) (_ : annot_o) (b : type) (_ : annot_o)
% | lambda (a b : type) | map (k : comparable_type) (v : type)
% | big_map (k : comparable_type) (v : type) | chain_id.
%
%
% Definition stack_type := Datatypes.list type.
% \end{lstlisting}
%
Instructions are defined in the \coqe{instruction} inductive type.
\coqe{instruction} is indexed by the type of the input and output
stacks. This indexing implies that only well-typed Michelson
instructions are representable in Mi-Cho-Coq.
\footnote{This is also the case in the OCaml Michelson interpreter via
  the use of GADTs.}
%
% Michelson has more than 80 instructions and in order to lower
% the number of constructors of the instruction abstract syntax tree, we
% distinguish instructions that do not contain subprograms as arguments
% from the others. Most of the instructions are in the first category.
% Moreover we distinguish instructions from instructions sequences in
% order to faciliate the splitting of contracts and be able to reason
% about fractions of the language.
% \todo{PUSH case}
% \todo{explain entry points, probably in the Michelson part}
%
% At the bottom, \coqe{opcode} instructions are instructions
% that do not contain code to be executed, which is the case for most
% instructions.
%
%
A full contract is a sequence of instructions respecting the calling
convention of Michelson mentioned above:\todo{smaller font for lstlisting}
\begin{lstlisting}
instruction ((pair params storage) :: nil) ((pair (list operation) storage) :: nil).
\end{lstlisting}
where \coqe{storage} is the type of the storage of the contract and
\coqe{params} the type of its parameter.
% new : entry-points, annotations, parsing, opam package

Implementation-wise, Coq's canonical structures are used to deal with the ad-hoc
polymorphism of some Michelson instructions (\emph{e.g} ?ADD? that can
add integers to ?timestamp? or ?mutez? or integers).
Coq's notations make contracts'appearance in Mi-Cho-Coq very close to actual contracts.

Also a lexer, parser and typechecker have been implemented, making it
possible to generate a Mi-Cho-Coq AST from a string representing a Michelson
contract.
Support for Michelson entry-points and annotations has been added.
\todo{explain entry points in michelson section}

\subsubsection{Semantics}
An interpreter for Michelson has been implemented as an evaluator
\coqe{eval}. Its simplified type is
^forall {A B : list type}, instruction A B -> nat -> stack A^
^-> M (stack B)^. Intuitively the interpreter takes a
sequence of instructions and an input stack and returns an output
stack. Since Michelson programs can explicitly fail the output stack
is embedded in an error monad ^M^. A \emph{fuel} argument is added to
enforce termination of the interpreter. This argument will decrease
every time in any recursive call to ^eval^. Note that the notion of
\emph{fuel} is different from \emph{gas}, which measures computation
costs.

% Michelson is statically typed: the elements of a stack, stacks,
% instructions and contracts have a type.
%\url{https://github.com/coq/opam-coq-archive/tree/master/extra-dev/packages/coq-mi-cho-coq/coq-mi-cho-coq.dev}

%\url{https://github.com/coq/opam-coq-archive/tree/master/extra-dev/packages/coq-albert/coq-albert.dev}

\subsection{Functional verification of Michelson smart contracts}

We have verified the functional correctness of Michelson contracts
with Mi-Cho-Coq, including complex ones such as a multisig contract
(cf. section 4 of~\cite{michocoq}) or a daily spending limit
contract~\footnote{\url{https://blog.nomadic-labs.com/formally-verifying-a-critical-smart-contract.html}}
used in the Cortez mobile wallet~\footnote{\url{https://gitlab.com/nomadic-labs/cortez-android}}.
Our correctness results are statements that
condition
successful runs of a contract  with the respect of a specification:

\begin{lstlisting}
Definition correct_smart_contract {A B : stack_type}
     (i : instruction A B) min_fuel spec : Prop :=
     forall (input : stack A) (output : stack B) fuel,
       fuel >= min_fuel input ->
       eval i fuel input = Return (stack B) output <->
         spec input output.
\end{lstlisting}

For example, for the voting contract described in
section~\ref{sec:michelson-vote}, the specification would be that
(preconditions) the amount sent is greater than or equal to 5 tez, the
chosen candidate is one the possible choices and that (postconditions)
the evaluation of the contract generates no operation, and modifies
only the votes count by incrementing by 1 the number of votes of the
chosen candidate.\todo{add pseudo-code}

In order to facilitate these functional proofs, a weakest precondition
calculus ^eval_precond^ is implemented. Its type is
^forall {fuel A B}, instruction A B -> (stack B -> Prop) ->^ ^(stack A -> Prop)^
that for an instruction and a postcondition (a predicate over the
output stack) returns the weakest precondition (a predicate over the
input stack).

The correctness of ^eval_precond^ has been proven:
\begin{lstlisting}
  Lemma eval_precond_correct {A B} (i : instruction A B) fuel st psi :
    eval_precond fuel i psi st <->
      match eval i fuel st with Failed _ _ => False | Return _ a => psi a end.
\end{lstlisting}
Intuitively, ^eval_precond_correct^ states that (left to right) the
computed by ^eval_precond^ is a precondition and that (right to left)
it is the weakest. This lemma is heavily used in the proofs of correctness.

% \begin{itemize}
% \item say it is an update of ~\cite{michocoq} and says that detailed
%   presentation of the proof of multisig is available.
% \item formalisation: some modifications
% \item probably no example?? refer to previous article for Mi-Cho-Coq
% \end{itemize}

\subsection{Optimiser}

A Michelson optimiser has been implemented in Mi-Cho-Coq. The purpose
of this optimiser is to simplify Michelson programs, thus reducing the
gas costs of executing them, without modifying their
semantics. % This has the benefits of making them more gas
% efficient.
Simplifications are basic at the moment: the goal is mainly
to clean programs generated from higher level languages by removing
useless stack manipulations instructions.
%Also simpler to analyse and slightly simpler to formally verify.

Optimisations are defined in one file, \verb+optimizer.v+. A first
step (\verb+dig0dug0+) removes useless instructions (?DROP 0?, ?DIG 0? and ?DUG 0?),
needless uses of ?DIP? (?DIP 0 i? is replaced with ?i?) and replaces
?DIG 1? and ?DUG 1? with ?SWAP?.
A second step (\verb+digndugn+) removes ?DIG n; DUG n? sequences.
A third step (\verb+swapswap+) removes ?SWAP ; SWAP? sequences.
A fourth step (\verb+push_drop+) removes ?PUSH ; DROP 1? and rewrites ?PUSH ; DROP n+1? into ?DROP n? (for ?n > 0?).
The \verb+visit_instruction+ function, similarly to the Visitor
Pattern~\cite{gamma1994design}, traverses a Michelson sequence
%\todo{explain why lambdas are not visited}
of instructions and applies one optimisation received as an argument.
Finally, the \verb+cleanup+ function applies the four optimisations
(in the order of their presentation above) to a sequence of
instructions.
%
% Note that the implementation does not contain a precisely defined
% rewriting system (with atomic rewriting rules, a congruence and
% transitive closure function, and proofs of termination and confluence)
% as this rewriting system would be fairly simple and uninteresting to
% us. \todo{is there a need for this sentence ; maybe too negative}
% (There would be only two critical pairs: ?DIG 0 ; DUG 0?
%   and ?DIG 1 ; DUG 1? that could be reduced by both step 1 or 2. For both pairs, it is
%   straightforward that they would eventually reduce to ?NOOP?.)
%   \todo{should be a footnote. fix issues with lst listings}

% Indeed, such a rewriting would be very the rewriting rules seemed
% trivial enough to convince us that the \verb+cleanup+ function is
% equivalent to a rewriting function where every
%
  We prove that the semantics of Michelson
  instructions are preserved by the optimisations. This is implemented
  in \verb+typed_optimizer.v+. The main theorem
  \verb+optimize_correct+ states that if an instruction sequence can
  by typechecked then its optimised version can also be typechecked
  with the same type; furthermore if the initial sequence runs
  successfully on some stack, then the optimised version runs also
  successfully on the same stack and they both return the same value.
\section{Albert}
\label{sec:albert}

Albert is an intermediate language for Tezos smart contracts with a
compiler written in Coq and that targets Mi-Cho-Coq.

We present in this section a high level overview of Albert's design and
features.
A more detailed presentation of Albert's syntax, typing rules and
semantics can be found in~\cite{albertWTSC20}.

\subsection{Design overview}
%\input{albert}
%As Michelson, Albert is statically typed.
The key aspect of Albert's design is the abstraction of Michelson
stacks by records with named fields. This gives two practical
benefits: unlike in Michelson, in Albert we do not need to care about
the order of the values and we can bind variables to names.
%Another important aspect of the langgure
Also, unlike Michelson where contracts can only contain one sequence
of instructions, it is possible in Albert to define multiple
functions, thus giving the possibility to implement libraries.
An important limitation of Albert is that resources are still being
tracked: variables are typed by a linear type system that enforces
that each value cannot be consumed twice. A {\bf dup} operation
duplicates resources that need to be consumed multiple times. A next
step would be to generate these operations in order to abstract data
consumption.

In a nutshell, each expression or instruction is typed by a pair of
record types whose labels are the variables touched by the instruction
or expression. The first record type describes the consumed values and
the second record type describes the produced values.
Thanks to the unification of variable names and record labels, records
in Albert generalise both the Michelson stack types and the Michelson
pair type.

Albert offers slightly higher-level types than Michelson: records
generalise Michelson's pairs and non-recursive variants generalise
Michelson's binary sum types as well as booleans and option types.
Variants offer two main operations to the user: constructing a variant
value using a constructor, and pattern-matching on a variant value.

The semantics of the Albert base language is defined in big-step
style.

We present in figure~\ref{fig:albert-example} the translation in Albert
of the voting contract described in
section~\ref{sec:michelson-vote}.
%  An example of a simple voting
% contract written in Albert is available here~\cite{albertVoting}. The
% user of the contract can only vote for a pre-defined set of options
% and must pay at least a certain amount of tokens for its vote to be
% considered.
The storage of the contract is a record with two fields: a
\verb+threshold+ that represents the minimum amout that must be
transferred, and an associative map, \verb+votes+, with strings as
keys (the options of the vote) and integers as values (the number of
votes for each associated key). The contract contains a \verb+vote+
function that checks that the parameter sent is one of the available
options, fails if not and otherwise updates the vote count. The main
function \verb+guarded_vote+ verifies that the amount of tokens sent
is high enough and if so, calls \verb+vote+.

\begin{figure}[htbp]
\begin{lstlisting}[language=albert,basicstyle=\footnotesize,numbers=left]
type storage_ty = { threshold : mutez; votes: map string nat }

def vote :
  { param : string ; store : storage_ty }  ->
  { operations : list operation ; store : storage_ty } =
      {votes = state; threshold = threshold } = store ;
      (state0, state1) = dup state;
      (param0, param1) = dup param;
      prevote_option = state0[param0];
      { res = prevote } = assert_some { opt = prevote_option };
      one = 1; postvote = prevote + one; postvote = Some postvote;
      final_state =  update state1 param1 postvote;
      store = {threshold = threshold; votes = final_state};
      operations = ([] : list operation)

def guarded_vote :
  { param : string ; store : storage_ty } ->
  { operations : list operation ; store : storage_ty } =
    (store0, store1) = dup store;
    threshold = store0.threshold;
    am = amount;
    ok = am >= threshold0;
    match ok with
        False f -> failwith "you are so cheap!"
      | True  t -> drop t;
          voting_parameters = { param = param ; store = store1 };
          vote voting_parameters
    end
\end{lstlisting}
\caption{\label{fig:albert-example}A voting contract, in Albert}
\end{figure}

\subsection{Implementation overview}
\label{sec:compiler}

Albert is formally specified with the Ott tool~\cite{ottLang} in a
modular way (one \verb+.ott+ file per fragment of the language). From
the Ott specification the Albert lexer and parser as
well as typing and semantic rules are generated in Coq. The type checker is a Coq
function that uses an error monad to deal with ill-typed programs.
There is no type inference, which should not be a problem since Albert
is supposed to be used as a compilation target.
%
%Big-step semantics of Albert have been implemented in Coq.

The Albert compiler is
written in Coq, as a function from the generated Albert grammar to the
Michelson syntax defined in Mi-Cho-Coq. The compiler is extracted to
OCaml code, which is more efficient and easier to use as a library.
Compilation of types, data and instructions are mostly straightforward, apart
from things related to records or variants. Records are translated
into nested pairs of values, variants into a nesting of sum types.
Projections of record fields are translated into a sequence of projections over
the relevant components of a pair. Pattern matching over variants are translated into
a nesting of \verb+IF_LEFT+ branchings. %Each branch of an Albert pattern-matching is
% translated in Michelson and inserted in the associated position of the Michelson
% \verb+IF_LEFT+ branchings tree.
%
A mapping from variable names to their positions in the stack exists
at every point in the program. This mapping is currently naive,
variables are ordered by the lexicographic order of their names. This
mapping is used in the translation of assignment instructions.

%\input{future}
% Albert, reasoning on a lifetime of a contract
%- Prove the correctness of the compiler.
% - Improve the architecture
% - Optimiser
% - Generation of dup operations, so that no need to track resources.

\section{Future and related work}
\label{sec:future}
\subsection{Towards stronger guarantees in the OCaml Michelson interpreter}

The OCaml implementation of Tezos contains an interpreter for
Michelson. This interpreter is implemented with GADTs in a
way that gives subject reduction for free: well-typed Michelson
programs cannot go wrong: with the calling convention,
we have the guarantee that well-typed programs will always be executed
with stacks of the right length with data of the right type.

Nonetheless, because of the limitations of the logic implemented in
OCaml, we are unable to reason directly about this interpreter. In
this section, we sketch two possible solutions to this problem that we
are currently exploring.

\subsubsection{From Coq to OCaml}

An obvious solution is to use Coq's extraction to produce OCaml code.
Coq's extraction mechanism relies on well-studied theoretical
grounds~\cite{Moh89a,Moh89b}, has been implemented for many
years~\cite{letouzey02,letouzey04} and has even been partially
certified~\cite{Glondu09,Glondu12a}. However, OCaml code produced by Coq's
extraction can contain \verb+Obj.magic+ to circumvent OCaml's less
expressive type-system. That is in particular the case for Coq code
that uses dependent types, such as Mi-Cho-Coq's interpreter.

Replacing the current OCaml Michelson interpreter with an extracted
version containing \verb+Obj.magic+ occurrences would be problematic.
The Michelson interpreter is part of the economic protocol which is
sandboxed by a small subset of OCaml modules that, for obvious safety
reasons, does not contain \verb+Obj.magic+. Lifting this restriction
would lower the guarantees provided by the OCaml type system and is
not a path we would like to take.

A better solution would be to implement a second Michelson interpreter
in Coq that would use simpler types so that its extraction is safe and
compiles in the economic protocol sandboxing environment.
In this second implementation, instructions' types would not be
indexed by the input and output stacks and, as a consequence, the
interpreter would be much more verbose, as all the ill-cases (e.g.
executing ?ADD? on a stack with only one element) would need to be
dealt with.
Proofs of equivalence between the typed interpreter and untyped
interpreter would be needed, as the reference implementation would be
the typed one, that uses dependent types, as the use of richer types
would make it safer.
It is also worth noting that any changes to the Michelson interpreter
have to be approved by a community vote as it is part of the economic
protocol that can be amended.

\subsubsection{From OCaml to Coq with coq-of-ocaml}

A reverse approach is to use the coq-of-ocaml~\cite{claret:phd} tool
to mechanically translate the OCaml Michelson interpreter into Coq
code and then to manually prove in Coq the equivalence of the
two interpreters.

coq-of-ocaml~\cite{claret:phd} is a work in progress effort to compile
OCaml code to Coq. As a simple example to
illustrate coq-of-ocaml, let us consider a polymorphic tree and a sum
function that sums the values of a tree of integers. The OCaml code
and its coq-of-ocaml translations are respectively in
figure~\ref{fig:tree-ocaml}) and figure~\ref{fig:tree-coq}. Notice
that in Coq, the type of the values stored in the trees appears in the
type of the tree, as the rich type system of Coq does not allow this
to remain implicit. It is possible to reason about the generated Coq
program. For example one could prove manually that the sum of a tree
containing only positive integers is positive
(~\ref{fig:tree-coq-proof}).
\begin{figure}[htbp]
\begin{lstlisting}[language=caml,basicstyle=\footnotesize,numbers=left]
type 'a tree =
| Leaf of 'a
| Node of 'a tree * 'a tree

let rec sum tree =
  match tree with
  | Leaf n -> n
  | Node (tree1, tree2) -> sum tree1 + sum tree2
\end{lstlisting}
\caption{\label{fig:tree-ocaml}Sum of a tree, in OCaml}
\end{figure}
\begin{figure}[htbp]
\begin{lstlisting}[language=coq,basicstyle=\footnotesize,numbers=left]
Inductive tree (a : Type) : Type :=
| Leaf : a -> tree a
| Node : (tree a) -> (tree a) -> tree a.

Arguments Leaf {_}.
Arguments Node {_}.

Fixpoint sum (tree : tree Z) : Z :=
  match tree with
  | Leaf n => n
  | Node tree1 tree2 => Z.add (sum tree1) (sum tree2)
  end.
\end{lstlisting}
\caption{\label{fig:tree-coq}Sum of a tree, in Coq}
\end{figure}
\begin{figure}[htbp]
\begin{lstlisting}[language=coq,basicstyle=\footnotesize,numbers=left]
Inductive pos : tree Z -> Prop :=
| PosLeaf : forall z, z > 0 -> pos (Leaf z)
| PosNode : forall t1, t2, pos t1 -> pos t2 -> pos (Node t1 t2).

Fixpoint positive_sum (t : tree Z) (H : pos t) : sum t > 0.
Proof.
  destruct H; simpl.
  - trivial.
  - assert (sum t1 > 0).
    now apply positive_sum.
    assert (sum t2 > 0).
    now apply positive_sum.
    lia.
Qed.
\end{lstlisting}
\caption{\label{fig:tree-coq-proof}The sum of a positive tree is positive, manual proof in Coq}
\end{figure}
The Michelson interpreter is much more complex than this simple
example and makes a heavy use of advanced OCaml features, such as
GADTs.
At the moment of writing, features such as
%GADTs, and also other features such as
side-effects, extensible types, objects and polymorphic variants, are
not supported yet by coq-of-ocaml.
Regarding GADTs, currently coq-of-ocaml translates them but in a way
that generates axioms for casting. Work is being done to try to have
an axiom-free translation.
%
% TODO say coq-of-ocaml's goal is to handle large code bases, such as Tezos economic protocol
% TODO shall we say that we've made experiences?
We have managed to translate the whole economic protocol of the
Babylon\footnote{Babylon is the codename of the second amendment to
  the economic protocol voted by the Tezos community. It has been
  superseded by Carthage in March 2020.} version of Tezos into Coq using
coq-of-ocaml with the caveat that Coq axioms are needed and that OCaml
annotations were added.
As the Michelson interpreter is part of the economic protocol, this
means we have a Coq translation of the OCaml interpreter. A next step
would be to prove the equivalence of the translated interpreter with
the Mi-Cho-Coq one.

%  This simple example shows the potential of coq-of-ocaml,
% coq-of-ocaml is very experimental and it is not yet possible to prove
% the equivalence Furthermore, it would
% % TODO big picture: benefits of having a mechanised translation from
% So far, only a subset of the functional component of OCaml is
% supported: records, types synonyms, mutually recursive types and plain
% modules. Side-effects as well as polymorphic variants, extensible
% types, objects and functors are not handled. Ongoing work is done to
% support first-class modules using dependent records in Coq, to deal
% with universes inconsistencies in the generated Coq code due to
% polymorphism, and to suppor GADTs, which is specially important since
% they are heavily used in the OCaml Michelson interpreter.

% \item coq-of-ocaml: \url{https://gitlab.com/nomadic-labs/mi-cho-coq/-/tree/dev/src/michocoq/of_ocaml}
%   - import of the typed IR of Michelson programs in OCaml
%   - link between the AST of Mi-Cho-Coq
%   - next prove the link between the OCaml interpreter and the Mi-Cho-Coq `eval` function

\subsection{Improvements to Mi-Cho-Coq and Albert}
\subsubsection{Mi-Cho-Coq}
Mi-Cho-Coq has axioms for domain specific opcodes that are harder to
implement such as instructions to query the blockchain environment for
relevant information (amount sent during the transaction, current
time, sender of the transaction,\dots), data serialisation and
cryptographic primitives. Work is being done to decrease the number of
axioms being used.
Another useful addition would be to extend the expressivity of the
framework by supporting mutual calls and calls to other contracts, as
well as reasoning on the lifetime of a contract.
Another issue that would need to be dealt with would be to implement
the gas model in Mi-Cho-Coq. A better or longer term solution would be
to replace gas accounting with static computation of execution costs
\emph{à la} Zen Protocol~\cite{zenprotocol_whitepaper}.
\todo{add adversarial model}
\subsubsection{Albert}
Albert is very much a work in progress. Next steps would be to have a
smarter implementation of the compiler that would produce optimised
code, as well as to prove the compiler correctness and meta-properties
of the Albert language such as subject reduction.
Longer term, we would like to implement a certified decompiler from
Michelson to Albert as well as a weakest-precondition calculus to
Albert in order to reason about Albert programs.

\subsection{Related work}
Despite the novelty of the field, many works regarding formal
verification of smart contracts have been published or announced.
The K framework has been used to formalise semantics of
smart-contracts languages for
Cardano\footnote{\url{https://github.com/kframework/plutus-core-semantics}},
Ethereum\footnote{\url{https://github.com/kframework/evm-semantics}}
and Tezos\footnote{\url{https://github.com/runtimeverification/michelson-semantics}}.
Concordium has developed the ConCert certification
framework~\cite{concertCPP20}, implemented in Coq, that permits to reason on the lifetime
of a contract.
Scilla~\cite{scilla2018,scilla2019}, the smart-contract language of Zilliqa has been formalised in
Coq as a shallow embedding. Its formalisation supports inter-contract
interaction and multiple calls over the lifetime of the contract.

Several high-level languages for Tezos smart contracts are being
developed~\cite{ligo,smartpy,archetype,juvix}. Programs in
the Archetype language\cite{archetype} can contain security properties
assertions or specifications expressed in logic formulae that are
translated to the Why3 platform and then verified by automatic provers
such as Alt-ergo, CVC or Z3.
Juvix~\cite{juvix} implements a variant of the quantitative type theory
which enables to track resources, similarly to Albert,
as well as to specify and verify smart contracts.

\todo{add biblio for rel work: qtt}


\bibliographystyle{splncs04}
\bibliography{../short_references}

% \appendix

% \section{Michelson Script of the Multisig Contract}
% \label{sec:multisig_appendix}
% \lstinputlisting{multisig.tz}

% \section{(REMOVE) Draft - ideas}
% \input{draft}

\end{document}

%% file: michelsonlang.tex
\lstdefinelanguage{michelson}{%
   columns=fullflexible,%
   basicstyle=\small\tt ,
   commentstyle=\slshape,%
   keywords={%
     \{,\},
     DROP, DUP, SWAP, PUSH, SOME, NONE, UNIT, IF_NONE,%
   PAIR, CAR, CDR, LEFT, RIGHT, IF_LEFT, IF_RIGHT, NIL,%
   CONS, IF_CONS, SIZE, EMPTY_SET, EMPTY_MAP, MAP, ITER,%
   MEM, GET, UPDATE, IF, LOOP, LOOP_LEFT, LAMBDA, EXEC,%
   DIP, FAILWITH, CAST, RENAME, CONCAT, SLICE, PACK,%
   UNPACK, ADD, SUB, MUL, EDIV, ABS, NEG, LSL, LSR,%
   OR, AND, XOR, NOT, COMPARE, EQ, NEQ, LT, GT, LE,%
   GE, SELF, CONTRACT, TRANSFER_TOKENS, SET_DELEGATE,%
   CREATE_ACCOUNT, CREATE_CONTRACT, CREATE_CONTRACT,%
   IMPLICIT_ACCOUNT, NOW, AMOUNT, BALANCE, CHECK_SIGNATURE,%
   BLAKE, SHA, SHA, HASH_KEY, STEPS_TO_QUOTA, SOURCE,%
   SENDER, ADDRESS,%
   CMPEQ,CMPNEQ,CMPLT,CMPGT,CMPLE,CMPGE,%
   IFEQ,IFNEQ,IFLT,IFGT,IFLE,IFGE,%
   IFCMPEQ,IFCMPNEQ,IFCMPLT,IFCMPGT,IFCMPLE,IFCMPGE,%
   FAIL,%
   ASSERT,%
   ASSERT_EQ,ASSERT_NEQ,ASSERT_LT,ASSERT_LE,ASSERT_GT,ASSERT_GE,%
   ASSERT_CMPEQ,ASSERT_CMPNEQ,ASSERT_CMPLT,ASSERT_CMPLE,ASSERT_CMPGT,ASSERT_CMPGE,%
   ASSERT_NONE,ASSERT_SOME,%
   ASSERT_LEFT,ASSERT_RIGHT,%
   UNPAIR,%
   },%
   alsoletter={'},
   upquote=true,
   keywordstyle={\bfseries\sffamily},%
   morekeywords=[2]{%
     key, unit, signature, option, list, set, operation, address,%
     contract, pair, or, lambda, big_map, map,%
     int, nat, string, bytes, mutez, bool, key_hash, %
     timestamp, 'a, 'b, 'S, 'p%
   },%
   keywordstyle=[2]{\bfseries\ttfamily},%
   classoffset=2,%
   morekeywords=[3]{%
     storage, parameter, code %
   },%
   keywordstyle=[3]{\bfseries},%
   sensitive,%
   comment=[l]\#,%
   morestring=[d]",%"
   literate={->}{{$\rightarrow{}$}}1%
}[keywords,comments,strings]%

\lstMakeShortInline[language=michelson,basicstyle=\ttfamily,keywordstyle=\bfseries\sffamily\small]!

%% file: albertlang.tex
\lstdefinelanguage{albert}{%
   columns=fullflexible,%
   basicstyle=\tt ,
   commentstyle=\slshape,%
   keywordstyle={\color{red}\sffamily},%
   keywords={%
     \{,\},
     type, def,
     dup, drop,
     car, cdr,
     match, with, end,
     assert_some, Some, None, True, False, Left, Right,
     for, map, loop_left, in, do, done,
     failwith,
     contract, address, implicit\_account, amount,
     noop
   },%
   alsoletter={'},
   keywordstyle={\color{purple}\sffamily},%
   morekeywords=[2]{%
     key, unit, signature, option, list, set, operation, address,%
     contract, pair, or, lambda, big_map, map,%
     int, nat, string, bytes, mutez, bool, key_hash, %
     timestamp%
   },%
   keywordstyle=[2]{\color{blue}\ttfamily},%
   classoffset=2,%
   morekeywords=[3]{%
     storage, parameter, code %
   },%
   keywordstyle=[3]{\bfseries},%
   sensitive,%
   comment=[l]\#,%
   morestring=[d]",%"
   literate={->}{{$\rightarrow$}}1%
}[keywords,comments,strings]%

\lstMakeShortInline[language=albert]?